\documentclass[preprints,review,accept,oneauthor,pdftex]{Definitions/mdpi} 

\firstpage{1} 
\pubvolume{8}
\issuenum{4}
\articlenumber{196}
\pubyear{2022}
\copyrightyear{2022}
\externaleditor{Academic Editors: Robert H. Bernstein, Bertrand Echenard} % For journal Automation, please change Academic Editor to "Communicated by"%to AE: please check if Academic Editor should be added
\datereceived{31 January 2022} 
\dateaccepted{9 March 2022} 
\datepublished{22 March 2022} 
\hreflink{https://doi.org/10.3390/universe8040196} % If needed use \linebreak
\Title{Search for Muon-to-Electron Conversion with the \mbox{COMET Experiment} $^\dagger$}

\TitleCitation{Search for Muon-to-Electron Conversion with the COMET Experiment}

\Author{Manabu Moritsu $^{1,2,\ddagger}$ \orcidA{} on behalf of the COMET Collaboration}

\AuthorNames{Manabu Moritsu}

\AuthorCitation{Moritsu, M., on behalf of the COMET Collaboration.}
\address{%
$^{1}$ \quad Department of Physics, Kyushu University, Fukuoka 819-0395, Japan;  moritsu.manabu@phys.kyushu-u.ac.jp\\
$^{2}$ \quad Institute of High Energy Physics, Chinese Academy of Sciences, Beijing 100049, China}%Please add post code. (or zip code in the US).

\conference {Proceedings of the 3rd J-PARC Symposium (J-PARC2019), Tsukuba, Japan,  23--27 September, 2019} 

\abstract{
Charged Lepton Flavor Violation is expected to be one of the most powerful tools to reveal physics beyond the Standard Model. 
The COMET experiment aims to search for the neutrinoless coherent transition of a muon into an electron in the field of a nucleus. 
Muon-to-electron conversion has never been observed, and can be, and would be, clear evidence of new physics if discovered. 
The experimental sensitivity of this process, defined as the ratio of the muon-to-electron conversion rate to the total muon capture rate, is expected to be significantly improved by a factor of 100 to 10,000 in the coming decade. 
The COMET experiment will take place at J-PARC with single event sensitivities of the orders of $10^{-15}$ and $10^{-17}$ in Phase-I and Phase-II, respectively. 
The ambitious goal of the COMET experiment is achieved by realizing a high-quality pulsed beam and an unprecedentedly powerful muon source together with an excellent detector apparatus that can tolerate a severe radiation environment. 
The construction of a new beam line, superconducting magnets, detectors and electronics is in progress towards the forthcoming Phase-I experiment. 
We present the experimental methods, sensitivity and backgrounds along with recent status and prospects. 
}

\begin{document}
%%%%%%%%%%%%%%%%%%%%%%%%%%%%%%%%%%%%%%%%%%

\section{Introduction}

Muons have played an important role in the history of particle physics since their discovery. 
The absence of the radiative muon decay, $\mu \to e \gamma$, during the early days~\cite{Hincks47} implied that the muon is a distinct elementary lepton rather than an excited electron. Today, muons are recognized as the charged lepton in the second {\it generation}. 
We know a muon decays into an electron with two neutrinos as $\mu \to e \nu_{\mu} \bar{\nu_{e}}$. 
The two neutrinos from a muon decay are unable to annihilate and thus should have different {\it flavor}; this is known as Lepton Flavor Conservation in the Standard Model (SM) of particle physics. 
Nowadays, muons are again attracting a great deal of attention to search for physics beyond the SM. 

This article describes an experiment to search for muon-to-electron ($\mu$-$e$) conversion, which is one of the lepton-flavor-violating processes in the charged lepton sector. 
The experimental sensitivity is expected to be significantly improved in the coming decade with planned experiments. 
An introduction to muon-to-electron conversion is shown below in this section. 
The experimental methods are described in Section~\ref{sec:experiment}, followed by the sensitivity and background estimation in Section~\ref{sec:sensitivity}. 
We discuss expandability and comparison with another experiment in Section~\ref{sec:discuss}, and finally describe the prospects in Section~\ref{sec:prospect}.

%-------------------------------------------------------------------------------
\subsection{Theoretical and Phenomenological Aspects}

Lepton flavor-violating muon decays are extremely suppressed in the SM even when neutrino oscillation effects are taken into account; e.g., the branching ratio of $\mu \to e \gamma$ can be calculated as~\cite{Petkov77, Marciano77, Lee77a, Lee77b}
 
\begin{equation} %------------------------------------------------------------
B(\mu \to e \gamma) = \frac{3\alpha}{32\pi} \left|\sum_{i=2,3}  U^{*}_{\mu i} U_{ei} \frac{\Delta m_{i1}^2}{{M_W}^2} \right| ^2 
\sim O(10^{-54}) , 
\label{eq:mueg}
\end{equation} %------------------------------------------------------------
where $U_{ij}$ are elements of the Pontecorvo-Maki-Nakagawa-Sakata matrix, and $\Delta m_{ij}^2$ represents the squared-mass differences between neutrinos. 
Using experimental values from neutrino oscillations, the branching ratio is found to be of the order of $10^{-54}$ due to the huge suppression factor of $(\Delta m_{ij}^2 / {M_W}^2)^2$. 
For coherent muon-to-electron conversion in the field of a nucleus, $\mu^- N \to e^- N$, where $N$ denotes a nucleus, the rate is found to be of a similar order. 
Note that the SM predictions are negligibly small from an experimental point of view. 
On the other hand, many well-motivated physics models beyond the SM predict sizable branching ratios within reach of experimental sensitivities in the near future~\cite{Kuno01, Merciano08, Mihara13, Bernstein13}. 
Hence, an observation of charged lepton flavor violation would provide clear evidence of new physics. 

From a simplified model-independent perspective, it is known that the $\mu^+ \to e^+ \gamma$ process is strongly sensitive to the photon-mediated dipole interaction, whereas the $\mu^- N \to e^- N$ process is sensitive to not only the dipole term but also contact terms~\cite{Gouvea09}; 
here, the contact terms indicate tree-level or box-diagram interactions mediated by massive unknown particles. 
Note that the situation is not so simple in reality; for a more detailed approach, see~\cite{Crivellin17}. 
Which terms are dominant depends highly on the models of new physics. 
Therefore, the ratio of strengths of $\mu^+ \to e^+ \gamma$ and $\mu^- N \to e^- N$ can be a powerful discriminator among models. 
For instance, supersymmetric (SUSY) loops favor the dipole terms and thus enhance $\mu^+ \to e^+ \gamma$, whereas models with extra U(1) symmetry, leptoquarks, or R-parity violating SUSY induce contact terms in a tree level and thus enhance $\mu^- N \to e^- N$. 

For $\mu$-$e$ conversion, since a muon interacts with a quark in protons or neutrons in a nucleus, the conversion rate depends on the target nucleus, and the dependence varies with different models~\cite{Kitano02, Cirigliano09, Davidson19, Davidson20}. 
Once $\mu$-$e$ conversion is discovered, a comparison of the rates for different target nuclei may help distinguish among models.

%-------------------------------------------------------------------------------
\subsection{Experimental Aspects}
\label{sec:expaspect}

Since $\mu$-$e$ conversion is an interaction with a nucleus, it is natural to use a negative muon beam so that the muon can be captured and undergo the lepton-flavor-violating interaction with the nucleus. 
Negative muons are stopped in a target material, form a muonic atom with a nucleus, and cascade down to the $1s$ ground state. 
Within the SM, the fate of a muon is either to decay in orbit (DIO) or undergo nuclear muon capture. 
Muon DIO is the $\mu^- \to e^- \nu_{\mu} \bar{\nu}_e$ decay of the bound-state muon, and nuclear muon capture is expressed as $\mu^- + (Z,A) \to \nu_{\mu} + (Z-1,A)$. 
For the $^{27}$Al nucleus case, the muon lifetime of 864~ns implies a fraction of 39\% for the former and 61\% for the latter~\cite{Suzuki87}. 
Then, if $\mu$-$e$ conversion happens, we would observe an electron without neutrinos as $\mu^- + (Z,A) \to e^- + (Z,A)$. 
Since the process is coherent, i.e., does not change the nuclear state, the observed electron has a specific energy, 
\begin{equation} %------------------------------------------------------------
E_{\mu e} = m_{\mu} - B_{\mu} - E_{\mathrm{rec}} , 
\label{eq:energy}
\end{equation} %------------------------------------------------------------
where $B_{\mu}$ and $E_{\mathrm{rec}}$ are the muon binding energy and the recoil energy of the nucleus, respectively. 
The converted electron energy, $E_{\mu e}$, is 104.97~MeV for the $^{27}$Al case. 
This single monoenergetic electron signature is an experimental feature of $\mu$-$e$ conversion. 

The branching ratio of $\mu$-$e$ conversion is defined by the ratio of the $\mu$-$e$ conversion rate to the total muon capture rate, namely, 
\begin{equation} %------------------------------------------------------------
B(\mu^- N \to e^- N) = \frac{\Gamma(\mu^- + N \to e^- + N)}{\Gamma(\mu^- + N \to \mathrm{all} \ \mathrm{captures})}. 
\label{eq:def}
\end{equation} %------------------------------------------------------------

The normalization to captures has an advantage in calculation since many details of the nuclear wavefunction cancel in the
ratio. 
The current experimental upper limit is $7 \times 10^{-13}$ (90\% C.L.) for a gold target obtained by the SINDRUM-II experiment at Paul Scherrer Institute (PSI)~\cite{Bertl06}.

%%%%%%%%%%%%%%%%%%%%%%%%%%%%%%%%%%%%%%%%%%%%%%%%%%%%%%%%%%%%%%%%%%%%%%%%
\section{Materials and Methods}
\label{sec:experiment}

\begin{figure}[H] %---------------------------------------------------------------------------------------------------------
\includegraphics[width=0.7\linewidth]{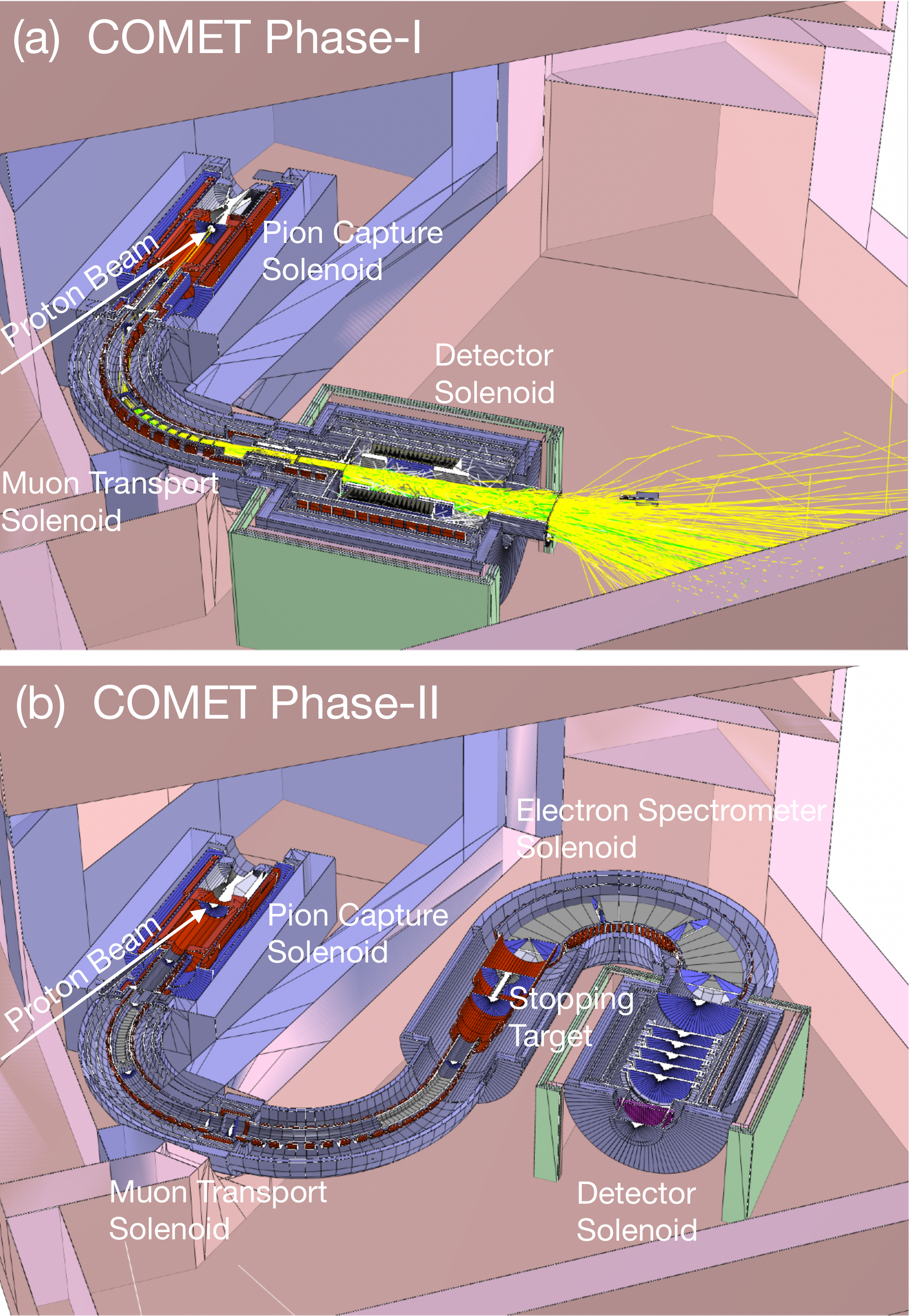}
\caption{Schematic layouts of the COMET experiment Phase-I (\textbf{a}) and Phase-II (\textbf{b}). Simulated particles for a single beam bunch are shown for Phase-I.}
\label{fig:layout}
\end{figure}

The COMET experiment~\cite{Kuno13, COMET20, Moritsu21} will search for $\mu$-$e$ conversion with a sensitivity of $O(10^{-17})$, which is four orders of magnitude better than the current limit. 
The experiment will take place at the Japan Proton Accelerator Research Complex (J-PARC). 
The construction of the dedicated beam line, muon source and detectors is in progress. 
Schematic layouts of the COMET experiment are shown in Figure~\ref{fig:layout}. 
The experiment adopts a staged approach, Phase-I and Phase-II, which will be described in Section~\ref{sec:staging}. 
In this section, we explain the conceptual design of the experiment; accelerator, beams and facility; and experimental apparatus.

 %---------------------------------------------------------------------------------------------------------

%-------------------------------------------------------------------------------
\subsection{Concept of the COMET Experiment}

\subsubsection{Highly-Intense Muon Source}
\label{sec:muonsource}

Going back in history, an original idea for modern $\mu$-$e$ conversion searches was proposed by Dzhilkibaev and Lobashev in 1989~\cite{MELC89} for the MELC experiment at INR, Russia, followed by the MECO experiment~\cite{MECO97} at BNL. 
Although neither of the experiments was realized, after 30 years, the concept has been inherited by the COMET and Mu2e~\cite{Pezzullo, Mu2e15} experiments today. 

One of the key concepts is an unprecedentedly powerful muon source. 
Instead of a conventional muon beam line using dipole and quadrupole magnets, all the sections related to pion production and transport use superconducting solenoid magnets. 
A proton beam impinges on a long target placed inside the production solenoid, and backward-generated low-energy pions are captured by a gradient solenoidal field which varies from 5~T at the target position to 3~T in the transport section. 
Note that the gradually-decreasing solenoidal field adiabatically translates part of the transverse momentum ($p_{\mathrm T}$) into the longitudinal component ($p_{\mathrm L}$), and make the particle trajectory more parallel, which helps increase the muon yield. 

Muons, as the decay product of pions, are efficiently delivered through a curved transport solenoid, where the center of their helical trajectories drifts vertically according to the particle charge ($q$) and momentum ($p$). 
The magnitude of the drift is given by 
\begin{eqnarray} %------------------------------------------------------------
D & = & \frac{1}{qB} \left( \frac{s}{R} \right) \frac{p_{\mathrm L}^2 + p_{\mathrm T}^2}{p_{\mathrm L}} \\
   & = & \frac{1}{qB} \left( \frac{s}{R} \right) \frac{p}{2} \left( \cos \theta + \frac{1}{\cos \theta} \right) , 
\label{eq:drift}
\end{eqnarray} %------------------------------------------------------------
where $B$ is the magnetic field along the axis; $s/R$ is a ratio of the path length to the curvature radius of the curved solenoid, i.e., the total bending angle; and $\theta$ is the pitch angle of the helical trajectory. 
In the COMET experiment, a compensating dipole field parallel to the drift direction is applied to keep the center of the helical trajectories of negative muons with momenta of 40 MeV/$c$ on the bending plane. 
The magnitude of the compensating field is given by 
\begin{eqnarray} %------------------------------------------------------------
B_{\mathrm{comp}} =  B \frac{D}{s} = \frac{1}{qR} \frac{p}{2} \left( \cos \theta + \frac{1}{\cos \theta} \right) , 
\label{eq:bcomp}
\end{eqnarray} %------------------------------------------------------------
for charged particles with momenta $p$ and pitch angles $\theta$. 
Positive or high-momentum muons are eliminated by a collimator placed after the curved solenoid. 

This type of pion-capture and muon-transport system has been demonstrated in the MuSIC beam line at Research Center for Nuclear Physics, Osaka University~\cite{Cook17}. 
The muon yield per beam power was found to be 1000 times better than the conventional method in existing facilities.

\subsubsection{Beam-Related Backgrounds and High-Quality Pulsed Beam}
\label{sec:beambg}

The past SINDRUM-II experiment~\cite{Bertl06} was performed by using a continuous beam at PSI.
Their results suggested a limitation on future experimental sensitivities from beam-related backgrounds caused by the continuous beam structure. 
The pulsed beam is therefore an essential requirement for next-generation $\mu$-$e$ conversion experiments. 
The COMET experiment utilizes a pulsed proton beam at J-PARC. 
In this subsection, we explain the beam-related backgrounds and measures to cope with them. 

Since the negative muon beam is generated by pion decay in flight, the muon beam is contaminated by pions, and its momentum spread is quite broad. 
Beam-related backgrounds are mainly caused by pions remaining in the muon beam. 
Radiative pion capture, $\pi^- + (Z,A) \to \gamma + (Z-1,A)$, followed by $\gamma \to e^+ + e^-$ conversion, is capable of generating a 105-MeV electron that mimics the $\mu$-$e$ conversion signal. 
Muon decays in flight may also emit a 105-MeV electron if the muon momentum is greater than 75~MeV/$c$. 
Antiprotons are negatively charged and therefore transported together with the negative muon beam. 
The antiprotons that annihilate in the stopping target could also produce \mbox{105-MeV electrons}.

\begin{figure}[H] %----------------------------------------------------------------------------
\includegraphics[width=0.8\linewidth]{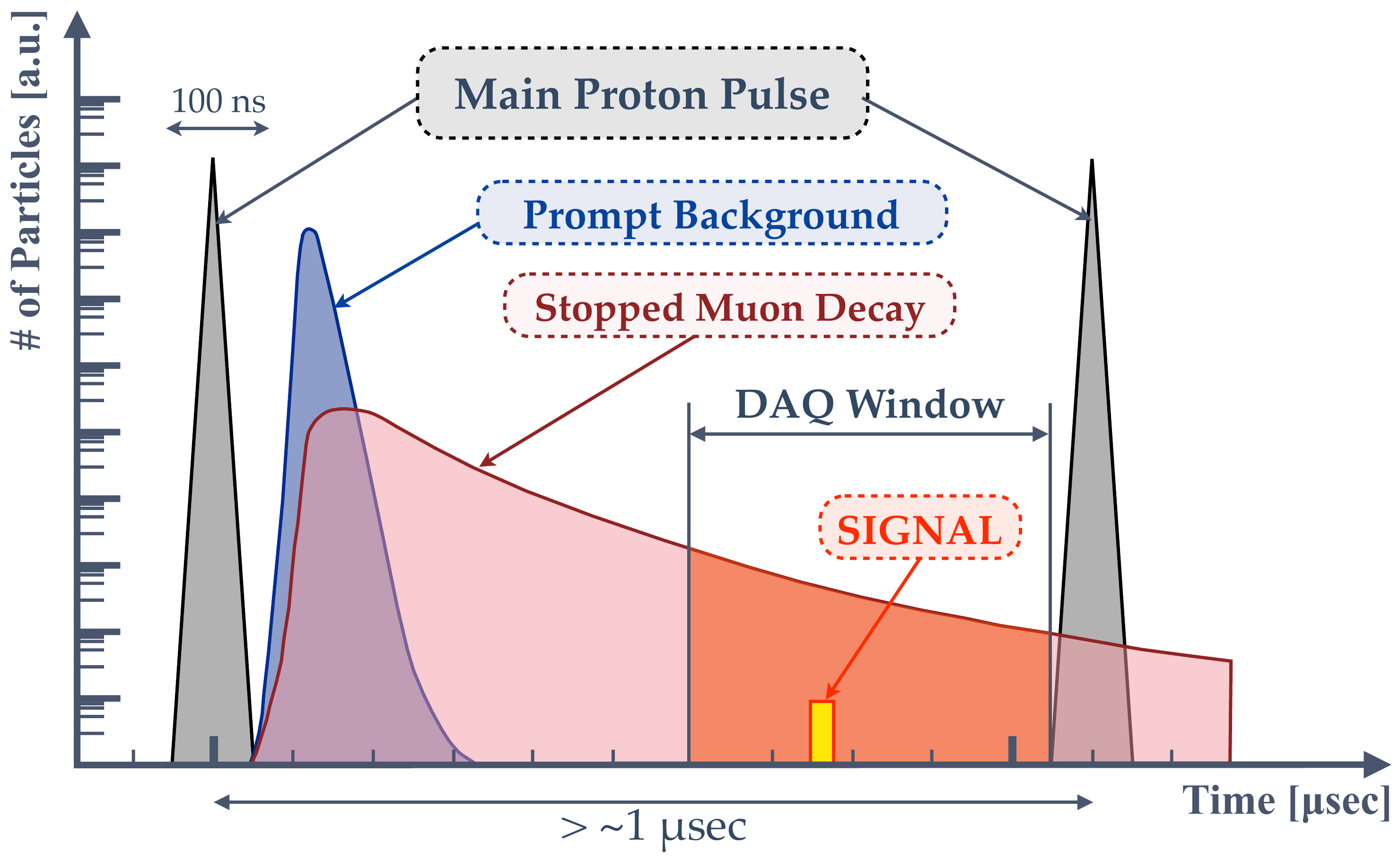}
\caption{Time relation among the proton beam pulses, prompt background, stopped muon decay, and a signal event in the data-taking window.}
\label{fig:extinct}
\end{figure}

Notice that the above beam-related backgrounds (except for delayed antiprotons) occur promptly, i.e., just after the primary proton beam timing, whereas the time distribution of $\mu$-$e$ conversion occurs according to the bound muon lifetime. 
As illustrated in Figure~\ref{fig:extinct}, the prompt backgrounds fade out in a few hundreds nanoseconds after the primary proton pulse. 
We therefore open the data-taking window from a few hundred nanoseconds after the proton pulse. 
Aluminum was selected as the target nucleus because the muon lifetime of muonic $^{27}$Al is 864~ns~\cite{Suzuki87}, long enough to perform the measurement. 
The data-taking window is closed before the next proton pulse, 1.2~$\upmu$s after the initial pulse. 

 %----------------------------------------------------------------------------
For the above scheme to work, cleanliness of the proton pulse structure is crucial. 
If there were leaked protons in between the primary proton pulses, it could produce prompt background in the data-taking window. 
The fraction of leaked protons in between the pulses, the so-called {\it extinction} factor, is required to be below $10^{-10}$ to achieve our sensitivity.

\subsubsection{Intrinsic Physics Background}

\begin{figure}[H] %----------------------------------------------------------------------------
\includegraphics[width=0.7\linewidth]{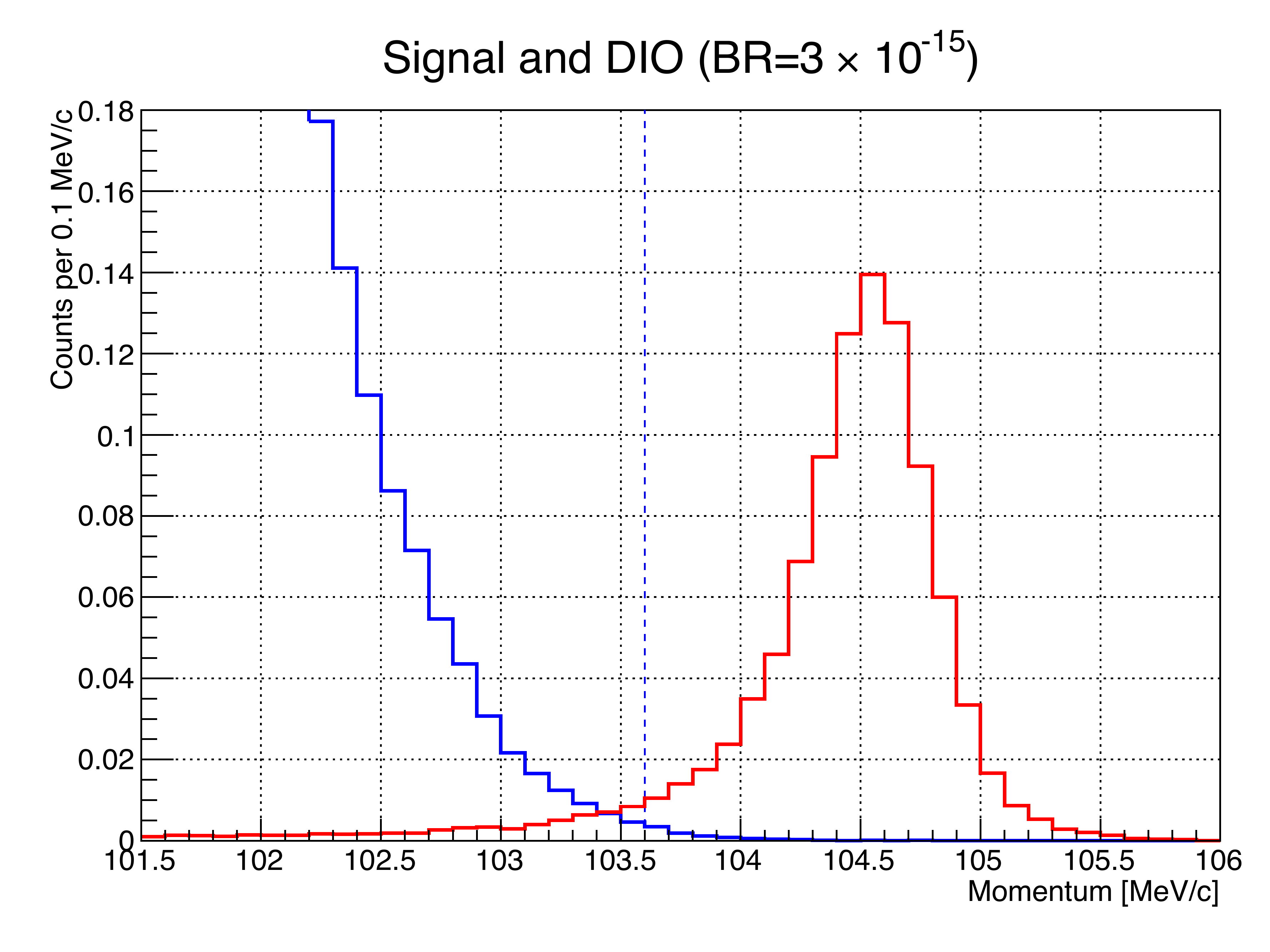}
\caption{Simulated electron momentum distributions of the $\mu$-$e$ conversion signal (red) and the decay-in-orbit events (blue), assuming a branching ratio of $3 \times 10^{-15}$ in COMET Phase-I. The vertical scale is normalized such that the integral of the signal is equal to one event.}
\label{fig:spectra}
\end{figure}

As described in Section~\ref{sec:expaspect}, 39\% of stopped muons decay in orbit in the muonic aluminum. 
The DIO electron energy distribution has a high-energy tail beyond the kinematic endpoint energy of normal muon three-body decay in free space, i.e., 52.8~MeV. 
It can be an intrinsic physics background. 
The energy tail extends in principle up to the $\mu$-$e$ conversion signal energy of 105~MeV, but falls steeply as $(E_{\mu e} - E_{\mathrm{DIO}})^5$~\cite{Czarnecki11}. 
Therefore, in order to distinguish the signal from the DIO background, a momentum resolution of 0.2\% is required for the electron detectors. 
Figure~\ref{fig:spectra} shows the expected momentum spectra for the signal and the DIO background.

It should be noted that the DIO tail endpoint energy depends on the atomic nucleus. 
Since light elements from boron ($Z=5$) to magnesium ($Z=12$) have higher endpoints than aluminum ($Z=13$), we should avoid using these materials in the detector region. 
 %----------------------------------------------------------------------------

\subsubsection{Cosmic-Ray Background}
\label{sec:crb}

Cosmic-ray induced backgrounds are not negligible in highly-sensitive $\mu$-$e$ conversion experiments. 
Cosmic-ray muons hitting the stopping target or the transport beam line may create 105-MeV/$c$ electrons, which mimic the signal of $\mu$-$e$ conversion. 
In addition, cosmic-ray muons, scattered at the stopping target or surrounding materials with a momentum near 105 MeV/$c$, are also hard to distinguish from the signal events. 
In order to suppress the cosmic-ray induced background, the detector region should be covered by veto counter arrays with a detection efficiency of 99.99\%. 
Note that cosmic-ray background events are proportional to the data-taking time, and therefore a shorter running time with higher beam intensity is preferable.

\subsubsection{Staging Scenario}
\label{sec:staging}

The COMET experiment adopts a staged approach to accomplish the physics goal in a timely manner. 
Experimental layouts of Phase-I and Phase-II are shown in Figure~\ref{fig:layout}. 

\begin{figure}[H] %----------------------------------------------------------------------------
\includegraphics[width=0.7\linewidth]{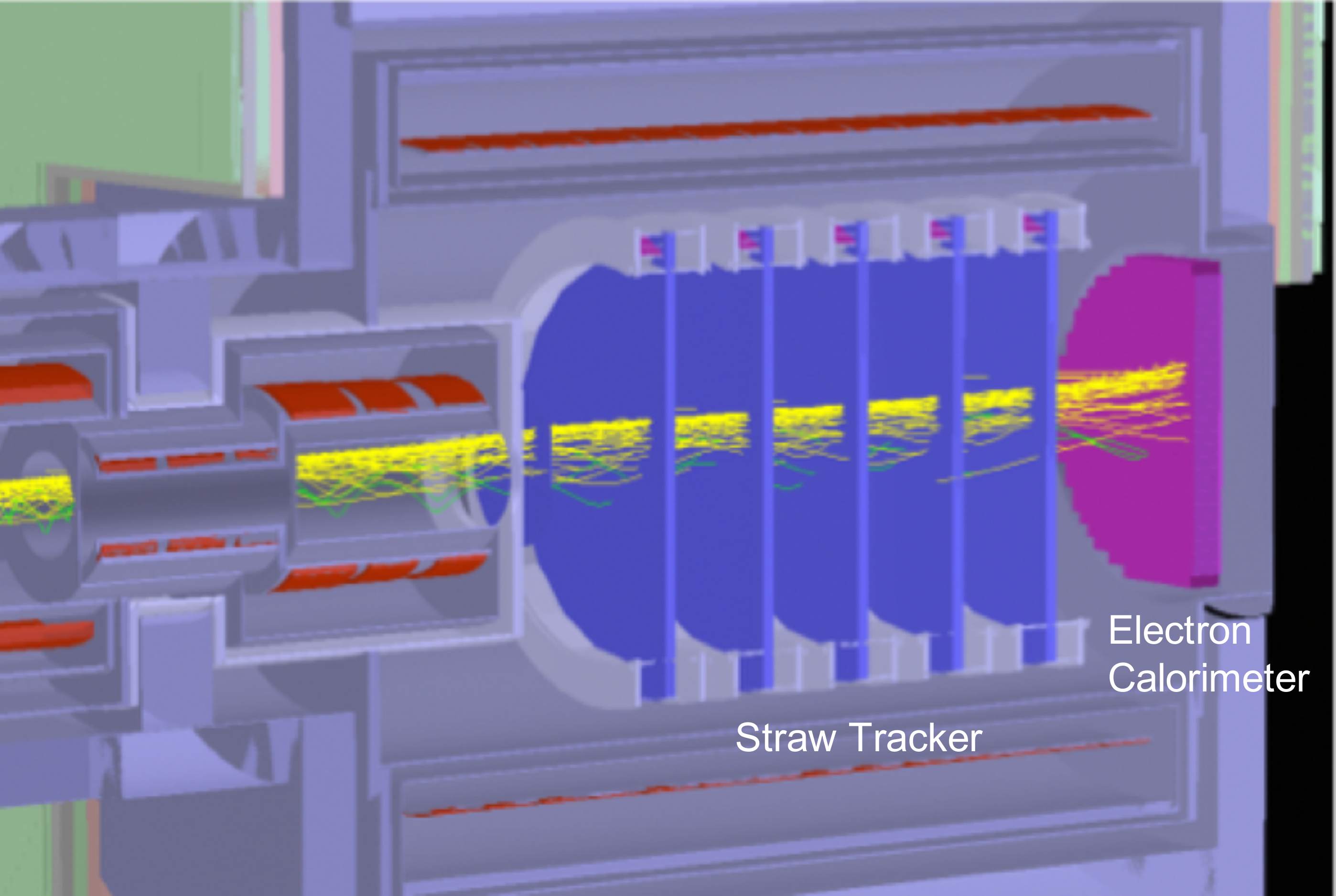}
\caption{Schematic layout of the StrECAL detector system for the muon beam measurement during Phase-I.}
\label{fig:strecal}
\end{figure}

\begin{figure}[H] %----------------------------------------------------------------------------
\includegraphics[width=0.8\linewidth]{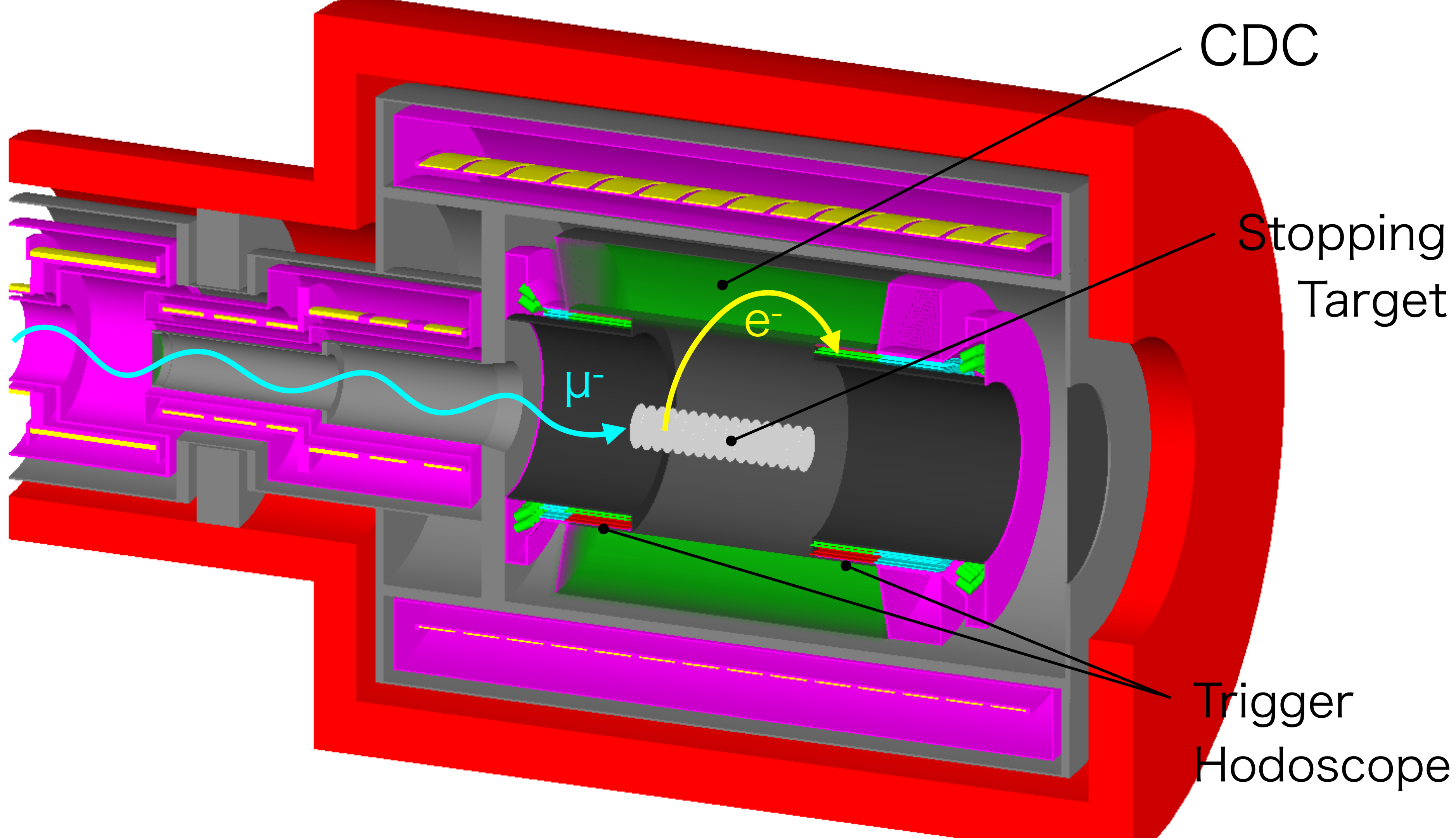}
\caption{Schematic layout of the CyDet system for the $\mu$-$e$ conversion search at Phase-I.}
\label{fig:cydet}
\end{figure}

In Phase-I, the Pion Capture Solenoid (CS), the Muon Transport Solenoid (TS), and the Detector Solenoid (DS) are constructed. 
The TS is constructed up to the end of the first $90^{\circ}$ bend. 
There are two goals in Phase-I: (1) muon beam measurement, and (2) $\mu$-$e$ conversion search with an intermediate sensitivity. 
(1) In the muon beam measurement, the beam quality at the end of the first $90^{\circ}$ bend position can be measured directly. 
This will help us understand the beam-related background more reliably based on real data instead of simulations. 
The muon beam measurement will be carried out with the {\it StrECAL} detector system, composed of straw-tube trackers and an electron calorimeter, as shown in Figure~\ref{fig:strecal}. 
The beam momentum, energy, and arrival time are measured. 
Details of the StrECAL detector system will be given in Section~\ref{sec:strecal}. 
(2) We will also conduct the $\mu$-$e$ conversion search with an intermediate sensitivity by using a primary proton beam power of 3.2~kW in Phase-I. 
The sensitivity goal is $3 \times 10^{-15}$, which is 100 times better than the current limit. 
This $\mu$-$e$ conversion search will be performed with the {\it CyDet} system, which consists of a cylindrical drift chamber (CDC) and cylindrical trigger hodoscopes (CTH) with a muon stopping target, as shown in Figure~\ref{fig:cydet}. 
The StrECAL system will be replaced by the CyDet system once the muon beam measurement is done. 
Muons stop in the aluminum target discs placed at the center of the DS. 
The transversely-emitted electrons from the muon stopping target are detected with the CDC and hit the CTH. 
Details of the CyDet system will be given later in Section~\ref{sec:cydet}.

 %----------------------------------------------------------------------------
\vspace{-6pt}
 %----------------------------------------------------------------------------

\begin{figure}[H] %----------------------------------------------------------------------------
\includegraphics[width=\linewidth]{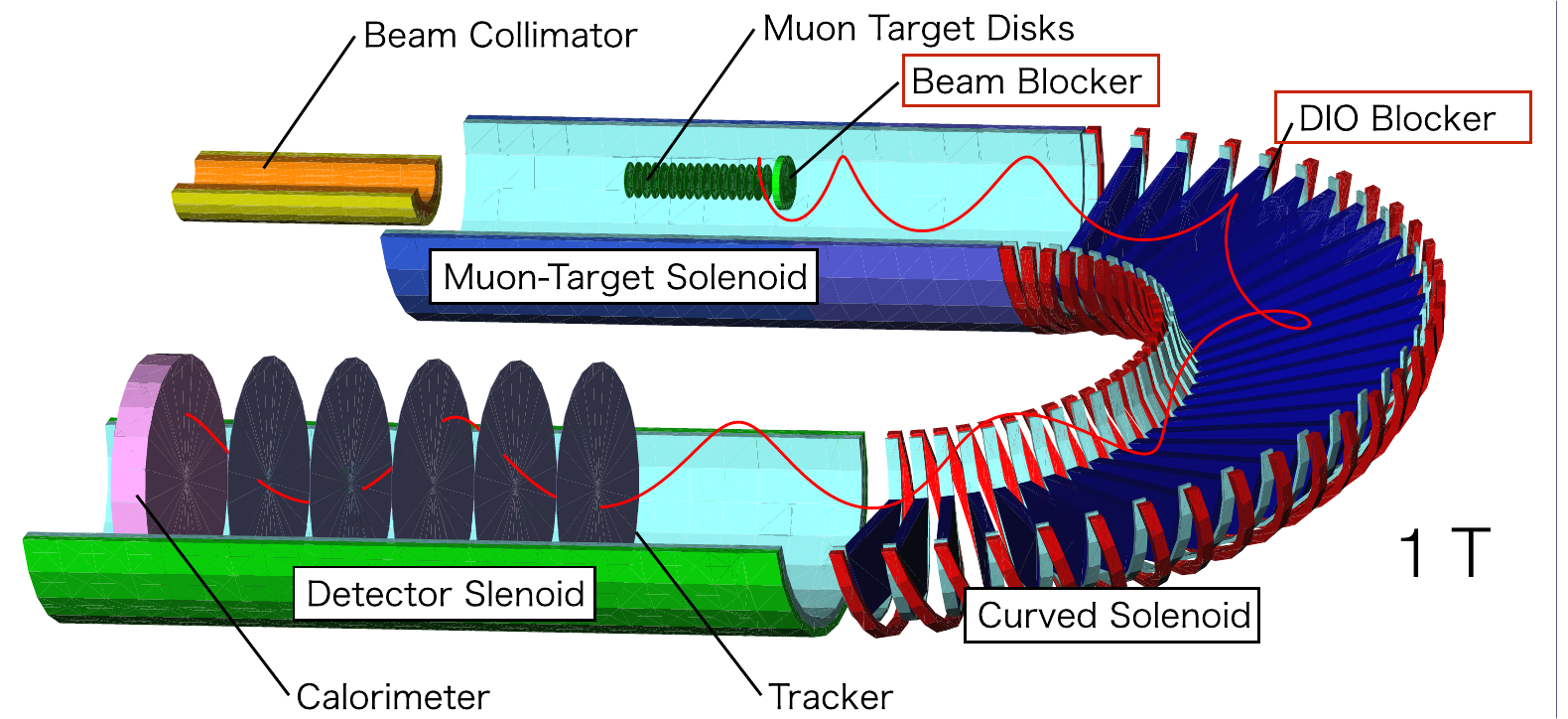}
\caption{Schematic layout of the downstream section of the Phase-II setup, which includes the muon stopping target, the Electron Spectrometer Solenoid, and the Detector Solenoid.}
\label{fig:spec-solenoid}
\end{figure}

In Phase-II, the beam power is increased to 56 kW. 
The TS is extended to the full $180^{\circ}$ bend, and the Electron Spectrometer Solenoid (ES) will be constructed. 
The ES is a $180^{\circ}$ bend curved solenoid installed at the downstream of the muon stopping section. 
It selects the charge and momentum of the emitted electrons with its curved solenoidal structure as shown in Figure~\ref{fig:spec-solenoid}. 
Low momentum electrons are eliminated by the DIO blocker installed at the bottom of the ES. 
Neutral particles can not directly reach the detector section. 
These help decrease the hit rate of the downstream detectors. 
The StrECAL system will be upgraded and used as detector in Phase-II. 
In this phase, we can improve the $\mu$-$e$ conversion sensitivity down to $2 \times 10^{-17}$, a 10,000 times better than the current limit.

 %----------------------------------------------------------------------------

%-------------------------------------------------------------------------------
\subsection{Accelerator, Beams and Facility}

\subsubsection{Accelerator and Proton Beam}
\label{sec:acc}

\begin{figure}[H] %----------------------------------------------------------------------------
\includegraphics[width=\linewidth]{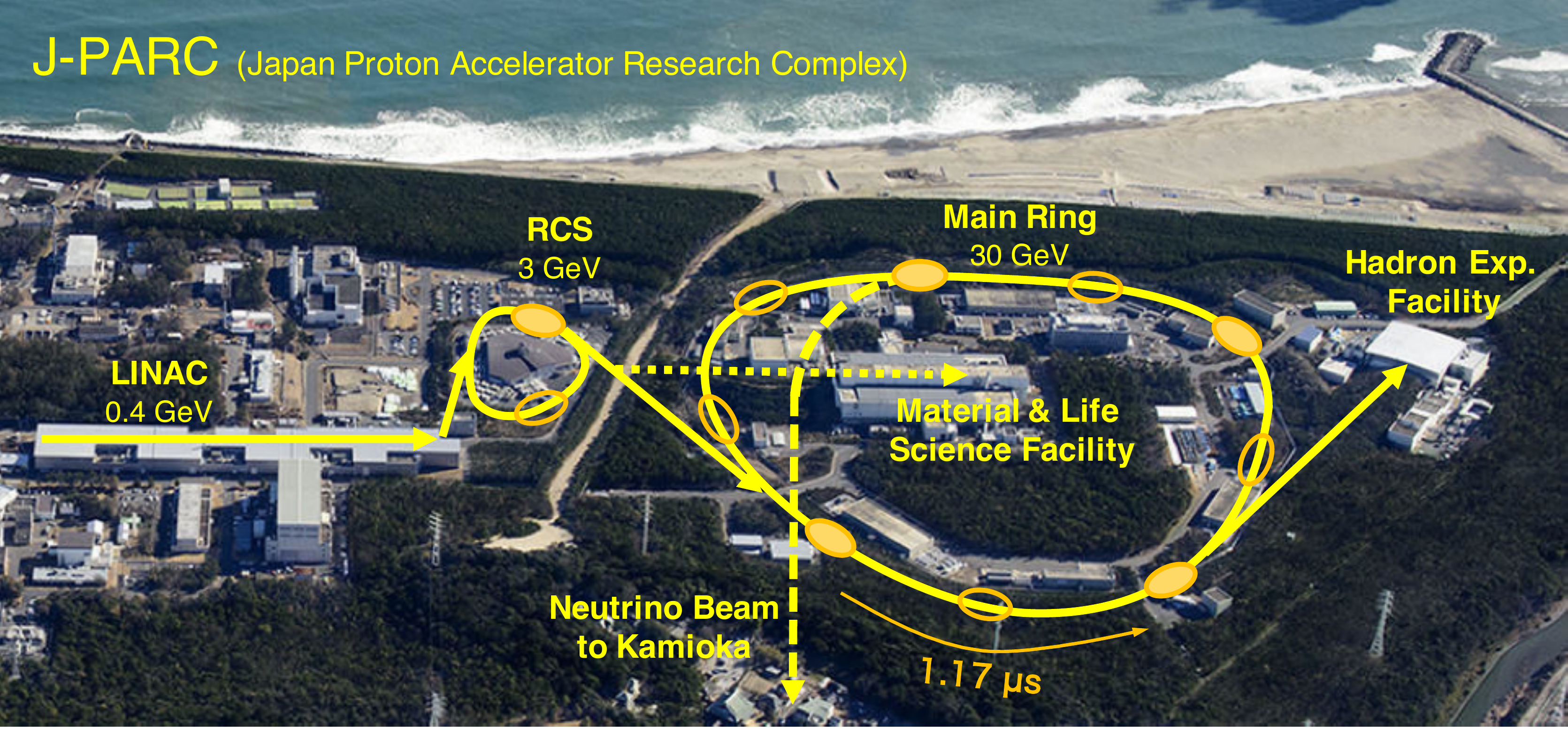}
\caption{A bird's eye view of the J-PARC accelerator facility. COMET takes place in the Hadron Experimental Facility. The orange ovals shown on RCS and Main Ring represent the proton bunch structure. The solid and open ovals indicate filled and empty proton bunches, respectively, in the COMET 8-GeV operation. (The photo was provided by the J-PARC Center).}
\label{fig:jparc-acc}
\end{figure}

The accelerator must provide a proton intensity as high as possible, and suppress beam-related backgrounds, as explained in Section~\ref{sec:beambg}. 
The COMET experiment requires a dedicated operation of the J-PARC accelerator. 
The proton beam energy is reduced from the normal Main Ring operation of 30 GeV to 8 GeV, which is sufficiently high to produce a large number of pions but low enough to minimize antiproton production. 
The beam power is adjusted to 3.2 kW in Phase-I, and will be upgraded up to 56 kW in Phase-II. 
We fill four  out of nine buckets of the Main Ring to realize the required pulse structure with an interval of 1.17 $\upmu$s as shown in Figure~\ref{fig:jparc-acc}. 
The proton beam is slowly extracted during \mbox{0.5 s}, keeping the bunch structure, and delivered through a new beam line to the COMET experimental hall. 
Most of the beam line components have already been constructed and will be ready to transport the beam in 2022.

 %----------------------------------------------------------------------------

A series of acceleration tests of 8-GeV proton beam and extinction measurements were carried out at J-PARC. 
The accelerator operation was successful with $7.3 \times 10^{12}$ protons per bunch, equivalent to the Phase-I design value. 
It was confirmed that the extinction factor satisfies the requirement of $10^{-10}$, and is expected to be reduced by improving the beam injection scheme to the Main Ring~\cite{Tomizawa19, Nishiguchi19}. 
In 2021, another test has been carried out to confirm the improved extinction scheme, and the result will be published soon~\cite{Noguchi21}. 

During physics data taking periods, online monitor devices installed in the proton beam line will measure the spill-by-spill extinction. 
In order to withstand the huge amount of radiation dose, e.g., $O(10^{19})$ total protons or higher, we have been developing wide-gap semiconductor detectors, such as chemical-vapor-deposition diamond~\cite{Fujii19}, silicon \mbox{carbide, etc.}

\subsubsection{Production Target and Superconducting Magnets}

A pion production target is enclosed in the superconducting solenoid magnet. 
A graphite target is used in Phase-I, and a tungsten target in Phase-II to increase the production rate. 
To mitigate severe radiation from the production target, the inner surface of the production region is protected with a radiation shield, and aluminum-stabilized NbTi wires are adopted as the superconducting coils~\cite{Yoshida13}. 

As described in Section~\ref{sec:muonsource}, the superconducting solenoid magnet system consists of the Pion Capture Solenoid, the Muon Transport Solenoid, and the Detector Solenoid as well as the Electron Spectrometer Solenoid for Phase-II, resulting in a total length of 15~m at Phase-I and 30~m at Phase-II. 
Manufacturing of the solenoid magnets~\cite{Yoshida15} including their cooling systems~\cite{Okamura20} is in progress, and will be completed in 2023.

%-------------------------------------------------------------------------------
\subsection{Experimental Apparatus}

\subsubsection{Straw-Tube Tracker and Electron Calorimeter}
\label{sec:strecal}

\begin{figure}[H] %----------------------------------------------------------------------------
\includegraphics[width=0.8\linewidth]{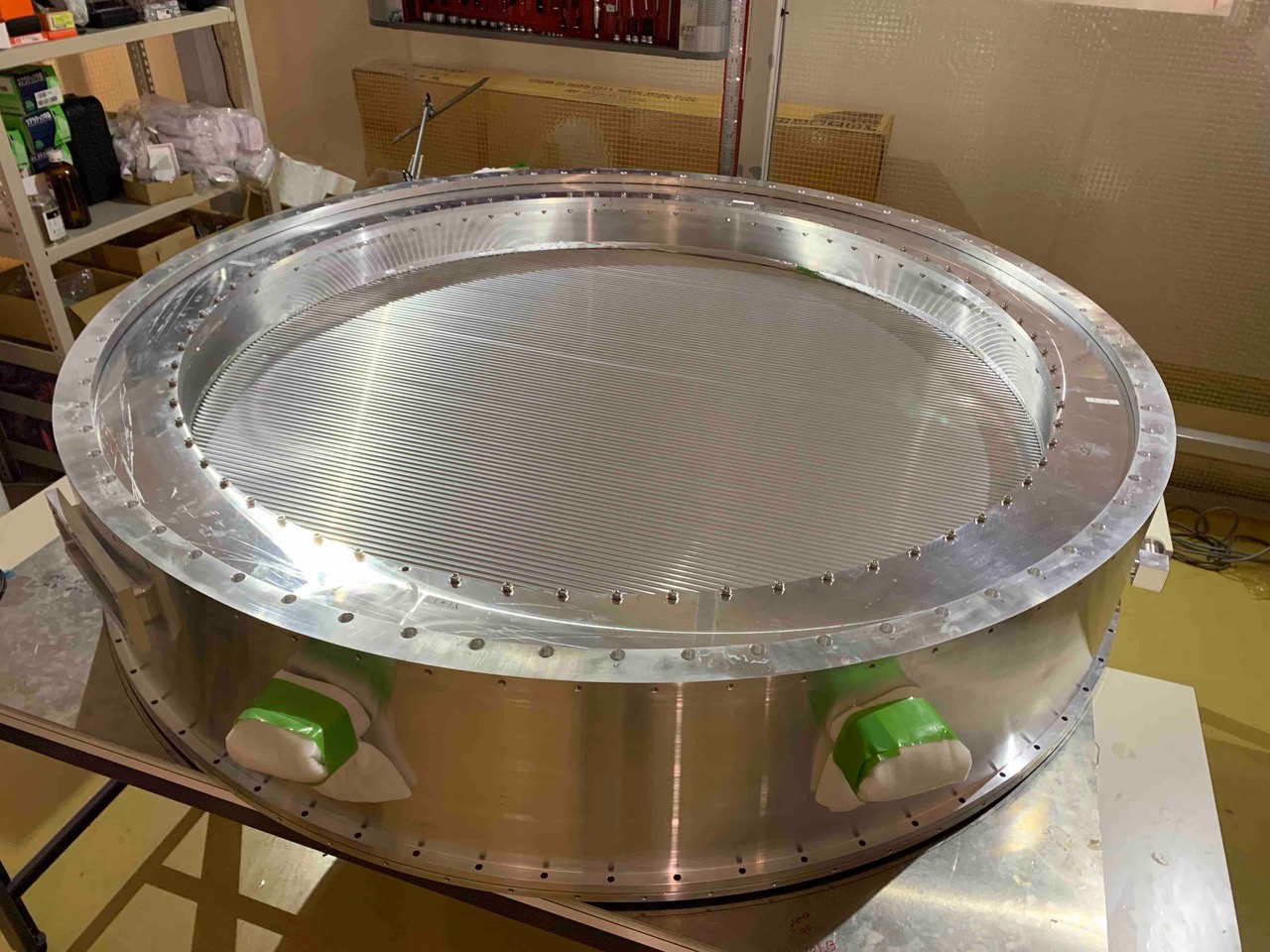}
\caption{Photograph of the straw tracker being assembled (first station).}
\label{fig:strawphoto}
\end{figure}

The StrECAL system is being developed for both the beam measurement program in Phase-I, and the $\mu$-$e$ conversion search in Phase-II (Figure~\ref{fig:strecal}). 
The thin-wall straw-tube planar tracker with extremely light materials operates inside a vacuum~\cite{Nishiguchi17, Volkov21}. 
The straw tubes are made of a 20-$\upmu$m thick Mylar foil with a 70-nm thick aluminum layer, and have a diameter of 9.8~mm and a length of more than 1~m. 
Straw tubes with even thinner walls of 12~$\upmu$m with a diameter of 5~mm are being developed for Phase-II, making use of ultrasonic welding techniques. 
The production of the 3400 straw tubes for Phase-I has been completed, and assembly of the tracker station is in progress as shown in Figure~\ref{fig:strawphoto}. 
Five stations of straw-tube trackers are used for tracking in the Detector Solenoid at 1~T. 
Each station consists of two horizontal and two vertical staggered pairs of planes.

 %----------------------------------------------------------------------------

The electron calorimeter (ECAL) is placed downstream of the straw trackers to measure the energy of electrons and provide a trigger signal. 
The ECAL adds redundancy to electron energy and hit position measurements, and provides initial parameters for precise tracking with the straw trackers. 
The ECAL consists of 1920 LYSO (Lu$_{2(1-x)}$Y$_{2x}$SiO$_{5}$) scintillating crystals with a cross section of 2 $\times$ 2 cm$^2$ and a depth of 12~cm corresponding to \mbox{10.5 radiation} lengths. 
The crystals are read out by avalanche photo diodes. 
A prototype beam test demonstrated an energy resolution better than 5\% and a fast decay time of 40~ns~\cite{Oishi18}. 

Signals from both the straw tracker and the ECAL are processed with front-end readout electronics, which contains preamplification, pulse shaping, discrimination, and digitization~\cite{Ueno19}. 
All the functionalities are controlled by FPGAs. 
The straw-tracker front-end boards are installed inside a gas manifold in vacuum.
To minimize the number of vacuum feedthrough connectors, we implemented a system to connect several boards with a daisy-chain of Gigabit Ethernet~\cite{Hamada21}.

\subsubsection{Cylindrical Detector System}
\label{sec:cydet}

\begin{figure}[H] %----------------------------------------------------------------------------
\includegraphics[width=0.8\linewidth]{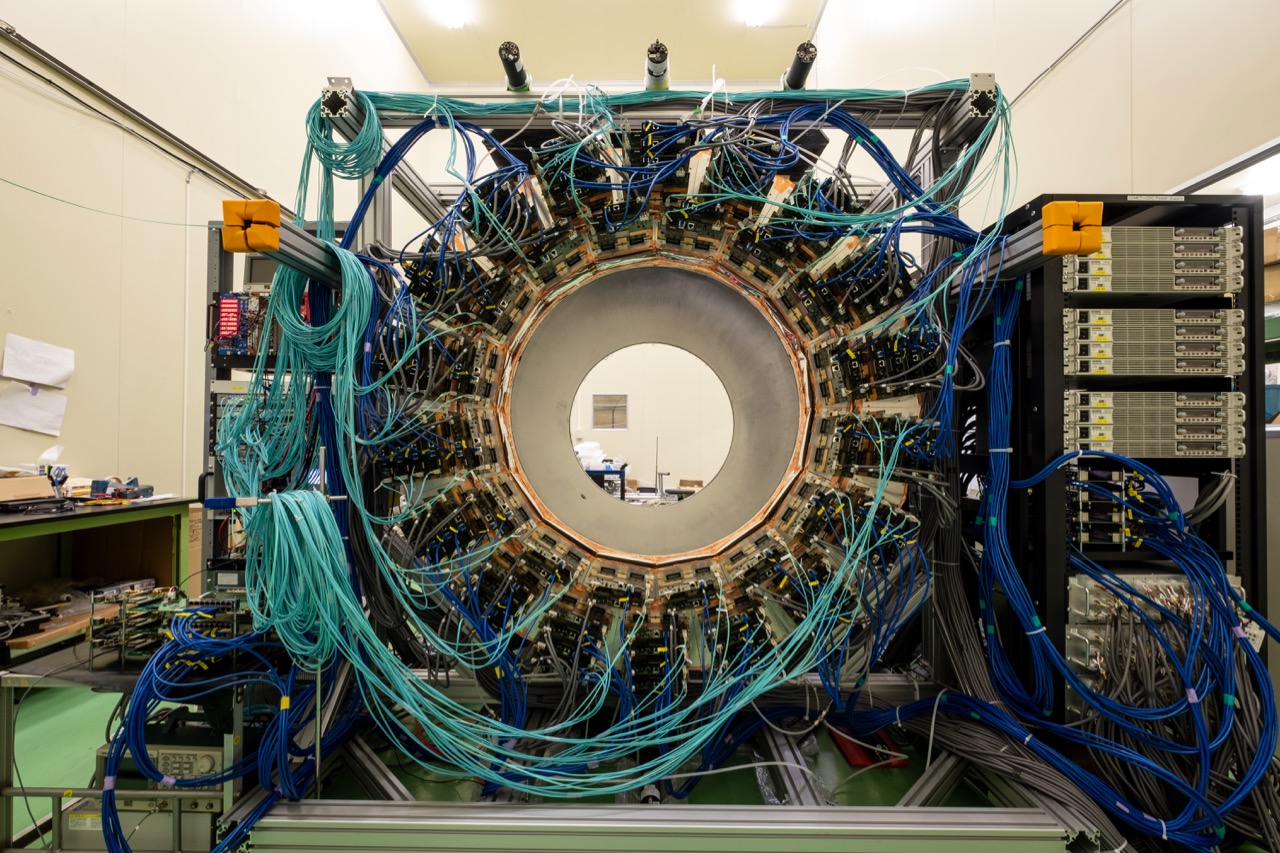}
\caption{Photograph of the CDC being tested with cosmic rays.}
\label{fig:cdcphoto}
\end{figure}

As introduced in Section~\ref{sec:staging}, the $\mu$-$e$ conversion search in Phase-I is conducted with the CyDet system (Figure~\ref{fig:cydet}). 
The momenta of transversely-emitted electrons are measured with the cylindrical drift chamber (CDC) in a magnetic field of 1 T. 
The CDC design was optimized to measure 105-MeV/$c$ electrons while suppressing unwanted low-energy particles. 
In order to achieve the required  momentum resolution of 0.2\%, the chamber is operated with a low-$Z$ gas mixture of He (90\%) and iC$_{4}$H$_{10}$ (10\%) to minimize multiple scattering effects. 
The inner cylinder of the CDC is made of carbon-fiber reinforced plastic with a thickness of 0.5~mm. 
The 4986 sense wires are made of gold-plated tungsten with a diameter of 25~$\upmu$m, while the 14562 field wires are made of unplated aluminum with a diameter of 126~$\upmu$m. 
An alternated all-stereo layer configuration is adopted to enhance the position resolution in the longitudinal direction. 
We tested a small CDC prototype with an electron beam, and demonstrated an average spatial resolution of 150 $\upmu$m with hit efficiency of 99\%~\cite{Wu21}. 
Construction of the CDC has already been completed and performance tests with cosmic rays are in progress~\cite{Moritsu19, Moritsu20} as shown in Figure~\ref{fig:cdcphoto}.

 %----------------------------------------------------------------------------

The first level trigger is issued by the cylindrical trigger hodoscope, which consists of a combination of plastic scintillation and acrylic Cherenkov counters. 
The high-level trigger is generated by using hit information of CDC at online level, resulting in an overall trigger rate down to about 10 kHz~\cite{Nakazawa21}. 

The muon stopping target is made of 17 aluminum discs with a thickness of 200~$\upmu$m, a radius of 100~mm, and a spacing of 50~mm. 
When the muonic atoms are formed on the muon stopping target, a cascade of X-rays is emitted as the muons drop down to the $1s$ state. 
The X-rays from muonic aluminum can be used to estimate the number of stopped muons, which leads to the denominator of the $\mu$-$e$ conversion branching ratio (Equation~(\ref{eq:def})). 
The major X-ray transition that we will use is the $2p \to 1s$ producing X-rays of 347~keV since it has the highest emission rate of 80\%
 with a relatively low background. 
The X-rays are counted with a high-efficiency Germanium detector installed in the downstream of the Detector Solenoid off the beam axis with a radiation shield. 

\subsubsection{Cosmic-Ray Veto Counters}

As described in Section~\ref{sec:crb}, the cosmic-ray veto (CRV) counter array is installed to cover all the Detector Solenoid. 
The main part of the CRV is composed of four layers of plastic scintillation counters read out by wave-length-shifting fibers and SiPMs. 
Note that the CRV counters suffer from a large neutron flux, especially in the upstream area. 
Since the hit rate of scintillators is predicted to be very high and the radiation damage to SiPM from neutrons is a concern, the upstream connection area between the Transport and Detector Solenoids is planned to be covered by glass resistive plate chambers.

\subsubsection{Trigger and Data Acquisition}

In the central trigger and timing system, the master trigger processor board collects the primary trigger, makes the final trigger decision, and provides the control clock to readout boards. 
To keep the commonality between the master trigger board and each frontend readout or trigger board connections using an optical serial link, a custom FPGA board has been developed, which is connected to the frontend boards. 
The high-level trigger system for CyDet is already described in Section~\ref{sec:cydet}.

The data acquisition system covers the data transfer from the frontend readout electronics to the data storage through an event builder. 
The system is being constructed with standard Ethernet network with the online software based on the MIDAS framework~\cite{Midas99}. 
The event-building throughput achieved is faster than 300~MiB/s and 1~GiB/s for the frontend and backend networks, respectively~\cite{Igarashi21}.

\subsubsection{Radiation Tolerance}

Radiation damage on frontend readout and trigger electronics is also an important issue in the experiment. 
The requirements for the radiation tolerance are a total dose of \mbox{1.0 kGy} and a neutron fluence of $1.0 \times 10^{12}$ cm$^{-2}$ for 1-MeV equivalent neutron in the severest case. 
We performed irradiation tests and selected electronics components that meet our requirements~\cite{Nakazawa19,Nakazawa20}.

\subsubsection{Offline Software}

To treat real and simulated data in the same way, we have developed an offline software framework called ICEDUST (Integrated Comet Experimental Data User Software Toolkit). 
A series of large-scale Monte Carlo simulations has been performed with each major software release to improve the experimental details. 
Since precise and clean track reconstruction is a challenging subject in the COMET experiment, several modern techniques are being developed, such as machine learning techniques for track reconstruction~\cite{Gillies18}, multi-turn track fitting algorithm~\cite{Zhang19}, and GPU-accelerated event reconstruction~\cite{Yeo21}.

%%%%%%%%%%%%%%%%%%%%%%%%%%%%%%%%%%%%%%%%%%%%%%%%%%%%%%%%%%%%%%%%%%%%%%%%
\section{Results---Sensitivity and Backgrounds}
\label{sec:sensitivity}

A summary of specifications of the COMET experiment is given in Table~\ref{tab:spec}. 
The single event sensitivity is given by 
\begin{equation}
B(\mu^- + \mathrm{Al} \to e^- + \mathrm{Al}) = \frac{1}{N_{\mu} \cdot f_{\mathrm{cap}} \cdot f_{\mathrm{gnd}} \cdot A_{\mu e}} , 
\label{eq:ses}
\end{equation}
where $N_{\mu}$ is the number of stopped muons, $f_{\mathrm{cap}}$ is the fraction which undergoes the nuclear muon capture, $f_{\mathrm{gnd}}$ is the fraction for which the final state of the nucleus is the ground state, and $A_{\mu e}$ is the overall acceptance including the efficiency of the detector system. 
In Phase-I (Phase-II), $1.5 \times 10^{16}$ ($1.6 \times 10^{18}$) stopped muons are accumulated in 150 days (\mbox{$\sim$1 year}), resulting in a single event sensitivity of $3.0 \times 10^{-15}$ ($2.0 \times 10^{-17}$). 

% The MDPI table float is called "specialtable" 
\begin{table}[H]
\setlength{\tabcolsep}{8.9mm}
\caption{Summary of the characteristics of Phase-I and Phase-II of the COMET experiment.}
\label{tab:spec}
\begin{tabular}{lll}
\toprule 
        & \bf{Phase-I}  & \bf{Phase-II} \\
\midrule 
Proton beam energy  &  8 GeV  &  8 GeV \\
Proton beam power  &  3.2 kW  &  56 kW \\
Total protons on target  &  $3.2 \times 10^{19}$  &  $1.0 \times 10^{21}$ \\
Total stopped muons  &  $1.5 \times 10^{16}$  &  $1.6 \times 10^{18}$ \\
Detector acceptance $\times$ efficiency  &  0.041  &  0.057 \\
DAQ time  &  150 days  &  260 days \\
Single event sensitivity  &  $3.0 \times 10^{-15}$  &  $2.0 \times 10^{-17}$ \\
Estimated background events &  0.032  &  0.66 \\
\bottomrule 
\end{tabular}
\end{table} %---------------------------------------------------------------------------------------------------------

Table~\ref{tab:bg} shows a summary of the estimated background events in the COMET experiment. 
In Phase-I, assuming a realistically-achievable extinction factor of $3 \times 10^{-11}$ and a time window from 700~ns after the beam bunch timing, the beam-related background can be suppressed to less than 0.01 events. 
Physics backgrounds such as DIO are estimated to be 0.01 events by selecting a momentum window from 103.6 to 106.0~MeV/$c$, while keeping the signal efficiency of 93\% (Figure~\ref{fig:spectra}). 
The estimated cosmic-ray background of less than 0.01 events is currently limited by the uncertainty on the simulation statistics. 
In total, the background events are estimated as 0.032 events in Phase-I. 

\begin{table}[H]
\setlength{\tabcolsep}{5.5mm} %---------------------------------------------------------------------------------------------------------
\caption{Summary of estimated background events for the COMET experiment~\cite{COMET20,Krikler16}.}
\label{tab:bg}
\begin{tabular}{llll}
\toprule 
\multirow{2}{*}{\bf{Type}} & \multirow{2}{*}{\bf{Background}} & \multicolumn{2}{c}{\bf{Estimated Events}} \\
              &                           &   \bf{Phase-I}        &  \bf{Phase-II} \\
\midrule 
\multirow{3}{*}{Beam}     &  Radiative pion capture &  0.0028        &  0.001  \\
             &  other prompt events    &  $<$0.0038  &  0.002 \\
             &  delayed antiproton       &    0.0012     &  0.296 \\
\midrule 
\multirow{3}{*}{Physics}  & Muon decay in orbit       &  0.01          &  0.068 \\
             & Radiative muon capture &  0.0019       &  $\sim$0 \\
             & Particle emission after muon capture &  $<$0.001  & ---  \\
\midrule 
Others   &  Cosmic rays               &  $\lesssim$0.01       &  0.294 \\
\midrule
Total     &                                    &  0.032           &  0.662 \\
\bottomrule 
\end{tabular}
\end{table}

The Phase-II sensitivity and background can be improved by tuning experimental parameters based on the Phase-I measurement results. 
The background is currently expected to be as small as 0.66 events~\cite{Krikler16}. 
In Phase-II, one dominant background source will be the cosmic-ray events because a larger area should be covered by veto counters including not only the Detector Solenoid but also the Electron Spectrometer Solenoid and the stopping target solenoid. 
The other dominant background source will be delayed-antiproton events, which in principle increase as the muon beam increases. 
These backgrounds should be refined using information from the Phase-I experiment.

 %---------------------------------------------------------------------------------------------------------

%%%%%%%%%%%%%%%%%%%%%%%%%%%%%%%%%%%%%%%%%%%%%%%%%%%%%%%%%%%%%%%%%%%%%%%%
\section{Discussions}
\label{sec:discuss}

\subsection{Further Improvements}

Further improvements towards $O(10^{-18})$ are being considered in the COMET Collaboration. 
Such improvements are likely to arise from refinements to the experimental design in Phase-II. 
We found that there is still room to optimize the position and size of the production target based on the latest estimates of the proton beam size. 
There is also room to increase the number of muon stopping target discs based on recent simulations. 
These improvements potentially increase the number of stopped muons by a factor of 3. 
An optimization on the height of the DIO blocker in the Electron Spectrometer Solenoid (Figure~\ref{fig:spec-solenoid}) has also been considered. 
The improvement potentially increases the signal acceptance by a factor of 2. 
If we assume 50\% longer running time than the nominal Phase-II and implement these optimizations, we can expect to improve the experimental sensitivity by one order of magnitude towards $O(10^{-18})$~\cite{COMET18}. 
It should be noted that the study is still preliminary and we have to carefully assess the backgrounds as well.

\subsection{Byproduct Experiments}

The COMET experimental setup is capable of accommodating byproduct experiments. 
Lepton-number-violating muon-to-positron conversion, $\mu^- + (Z,A) \to e^+ + (Z-2,A)$ (hereafter $\mu^{-}$-$e^{+}$ conversion), is a promising candidate; see also an article in this issue by Lee and MacKenzie~\cite{MJLee}. 
It could provide insight into the Majorana property of neutrinos. 
In the $\mu^{-}$-$e^{+}$ conversion search, the radiative muon capture (RMC) followed by $\gamma \to e^+ e^-$ could be the most significant background. 
The kinematic endpoint of the positron energy from RMC is higher than the $\mu^{-}$-$e^{+}$ conversion signal energy in case of the Al target. 
Therefore, we might need to consider other isotopes, e.g., $^{40}$Ca, $^{32}$S, or $^{48}$Ti, as an alternative stopping target~\cite{Yeo17}. 
By using an optimal target, the sensitivity is expected to be improved by four orders of magnitude for a number of stopped muons equivalent to Phase-II. 
It should be noted that we have to flip the polarity of the compensating dipole field of the Electron Spectrometer Solenoid so that positive particles are accepted. 

Another candidate is $\mu^- + e^- \to e^- + e^-$ in a muonic atom~\cite{Koike10, Uesaka16, Uesaka18}. 
The Coulomb attraction from the nucleus in a heavy muonic atom leads to a significant enhancement in its rate. 
Since the experimental signature is two nearly back-to-back electrons with momenta close to 50 MeV/$c$, we need to decrease the magnetic field of the Detector Solenoid in Phase-I, making this measurement presumably difficult to conduct in Phase-II. 

In addition, $\mu^- \to e^- + a$ in a muonic atom is being considered, where $a$ can be an axion-like particle, familon, or majoron with a lepton-flavor-violating coupling to leptons. 
Since the emission of the neutral particle $a$ distorts the DIO electron spectrum, we need to detect the subtle distortion with high precision~\cite{Wu20}.

\subsection{Comparison with the Mu2e Experiment}

COMET and Mu2e are competitive experiments to search for $\mu$-$e$ conversion. 
The final experimental sensitivity is at a similar level of $2 \times 10^{-17}$. 
Here, we briefly summarize differences between COMET (Phase-II) and Mu2e. 

\begin{itemize} %----------

\item Their strategies to realize the beam extinction are different. 
COMET takes place at J-PARC, whereas Mu2e is performed at FNAL. 
As described in Section~\ref{sec:acc}, COMET utilizes four of the existing nine buckets of the Main Ring to realize the required pulse timing structure. 
Therefore, in terms of the extinction, residual protons in the empty buckets should be mainly taken care of. 
On the other hand, Mu2e manipulates the beam bunch structure from two to four bunches during beam delivery. 
Therefore, straying protons in between bunches need to be eliminated with an additional extinction device, such as a dipole magnet with time-varying field (AC dipole) that sweeps out-of-time protons out of the beam. 

\item The shapes of the curved Transport Solenoids (TS) are different. 
Mu2e adopted a ``S'' shape TS with $90^{\circ}$ bend followed by $-90^{\circ}$ bend to compensate the vertical drift in Equation~(\ref{eq:drift}). 
As described in Section~\ref{sec:muonsource}, COMET adopted a ``C'' shape TS with $180^{\circ}$ bend, compensating the vertical drift by the dipole field embedded in TS. 
Since the magnitude of the vertical drift is proportional to the bending angle, we expect good separation of particle charge and momentum. 

\item The shapes of the Electron Spectrometer Solenoids are different. 
Mu2e adopted a conventional straight-shape solenoid which accommodates straw-tube trackers and electron calorimeters. 
The detectors have a hole in the central axis region to avoid hits from low-momentum particles. 
As shown in Figure~\ref{fig:layout}b, COMET adopted another ``C'' shape curved solenoid with $180^{\circ}$ bend. 
Unwanted low-momentum particles can be eliminated by the vertical drift of the curved solenoid with a compensation dipole field before reaching the tracker. 

\item The primary beam intensities are different.  
COMET utilizes a proton beam with \mbox{56 kW} and a data taking period of about one year; whereas Mu2e uses a 8-kW beam with about three years of data taking. 
Higher beam power shortens the data taking time although we need to handle more severe radiation environment and a higher particle rate. 
It should be noted that the data taking time does not take into account time sharing with other experiments in the same facility. 
COMET needs to share the J-PARC machine time with other neutrino or hadron experiments, while Mu2e can run in parallel with a neutrino experiment in FNAL. 

\item COMET adopted a two-staged approach as described in Section~\ref{sec:staging}, while Mu2e plans on a single stage. 
Since the final goal is to improve the sensitivity by a factor 10,000 with respect to the current limit, COMET has chosen to climb step by step even if it costs time. 
The Phase-I results will improve the detailed design of Phase-II to mitigate the risks. 

\end{itemize} %----------

We hope both experiments should be competitive but collaborative in some aspects, e.g., critical components for radiation damage, accelerator operation expertise, etc.

%%%%%%%%%%%%%%%%%%%%%%%%%%%%%%%%%%%%%%%%%%%%%%%%%%%%%%%%%%%%%%%%%%%%%%%%
\section{Prospects}
\label{sec:prospect}

\begin{figure}[H] %----------------------------------------------------------------------------
\includegraphics[width=\linewidth]{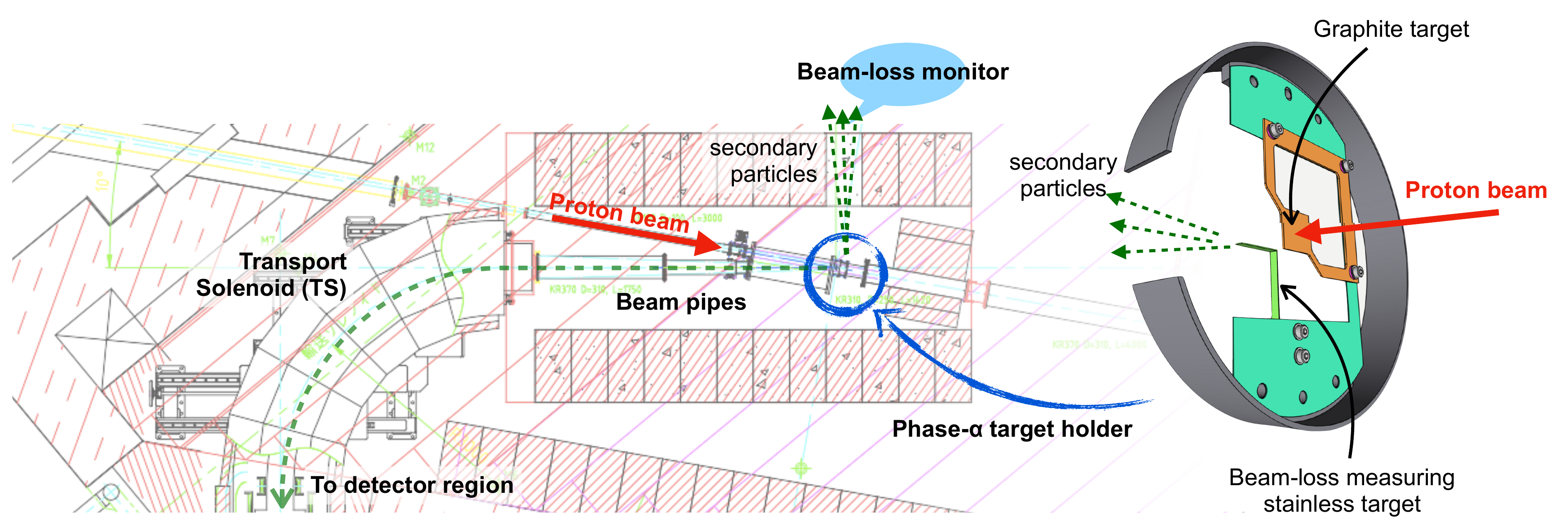}
\caption{Setup for the first beam commissioning phase (Phase-$\alpha$). A picture of the production target holder is also shown on the right.}
\label{fig:phase-a}
\end{figure}

We plan to conduct the first beam transport to the COMET experimental facility in the beginning of 2023. 
The proton beam commissioning and a muon beam yield and profile measurement will be performed with a simplified setup without the Pion Capture Solenoid: this is called {\it Phase-$\alpha$}~\cite{Phase-a}. 
Figure~\ref{fig:phase-a} shows the setup around the production target in Phase-$\alpha$. 
We use a stainless steel L-shape target to measure the horizontal and vertical positions of the proton beam. 
Secondary particles from the proton beam interaction with the L-shape target will be measured by a beam-loss monitor through a side hole in a radiation shield. 
We also use a graphite plate target with a thickness of 1 mm for muon yield and profile measurements. 
Since the Pion Capture Solenoid will not be installed at that time, muons that accidentally come into the Transport Solenoid (TS) are delivered to the downstream detector region. 
We expect about 5~kHz of negative muons at the exit of the TS with 200-W beam power. 
Since the Detector Solenoid will not be installed at that time, we plan to install scintillating fiber and hodoscope counters with a drift chamber at the downstream of the TS to measure the beam profile. 
A range counter composed of several layers of scintillators and absorbers will be also installed. 
The energy deposit, range, and time of flight information are used for particle identification. 
The Phase-$\alpha$ measurement will provide the muon yield and demonstrate the charge and momentum separation in \mbox{the TS}. 

After Phase-$\alpha$, we will install the Pion Capture and Detector Solenoids and move into Phase-I as described in Section~\ref{sec:staging}. 
The Phase-I experiment will start in 2024.

 %----------------------------------------------------------------------------

%%%%%%%%%%%%%%%%%%%%%%%%%%%%%%%%%%%%%%%%%%%%%%%%%%%%%%%%%%%%%%%%%%%%%%%%
\section{Conclusions}

The COMET experiment searching for $\mu$-$e$ conversion at J-PARC was described. 
The dedicated accelerator operation for the 8-GeV proton is ready, and construction of the beam line, superconducting magnets and detector system is progressing. 
The first beam commissioning is about to start, and the Phase-I experiment will follow. 
Phase-I encompasses both the muon beam measurement and $\mu$-$e$ conversion search with an intermediate sensitivity of $O(10^{-15})$ which is 100 times better than the current limit. 
The muon beam measurement will provide us a deeper understanding of the background, which will be incorporated in Phase-II. 
By increasing the beam power and improving the experimental apparatus, the Phase-II experiment will further improve the sensitivity, reaching $O(10^{-17})$. 
A possible discovery would open the door leading us to physics beyond the Standard Model.

%%%%%%%%%%%%%%%%%%%%%%%%%%%%%%%%%%%%%%%%%%
\vspace{6pt} 

%%%%%%%%%%%%%%%%%%%%%%%%%%%%%%%%%%%%%%%%%%
%% optional
%\supplementary{The following are available online at \linksupplementary{s1}, Figure S1: title, Table S1: title, Video S1: title.}


%%%%%%%%%%%%%%%%%%%%%%%%%%%%%%%%%%%%%%%%%%
% \authorcontributions{For research articles with several authors, a short paragraph specifying their individual contributions must be provided. The following statements should be used ``Conceptualization, X.X. and Y.Y.; methodology, X.X.; software, X.X.; validation, X.X., Y.Y. and Z.Z.; formal analysis, X.X.; investigation, X.X.; resources, X.X.; data curation, X.X.; writing---original draft preparation, X.X.; writing---review and editing, X.X.; visualization, X.X.; supervision, X.X.; project administration, X.X.; funding acquisition, Y.Y. All authors have read and agreed to the published version of the manuscript.'', please turn to the  \href{http://img.mdpi.org/data/contributor-role-instruction.pdf}{CRediT taxonomy} for the term explanation. Authorship must be limited to those who have contributed substantially to the work~reported.}

 \funding{The author was supported in part by the Japan Society of the Promotion of Science (JSPS) KAKENHI Grant No.~19K14747, and the Chinese Academy of Sciences President's International Fellowship Initiative.}%Please add: ``This research received no external funding'' or ``This research was funded by NAME OF FUNDER grant number XXX.'' and  and ``The APC was funded by XXX''. Check carefully that the details given are accurate and use the standard spelling of funding agency names at \url{https://search.crossref.org/funding}, any errors may affect your future funding.}

\acknowledgments{The author acknowledges strong support from the COMET Collaboration. This article was written on behalf of the COMET Collaboration. We acknowledge support from JSPS, Japan; Belarus; NSFC, China; IHEP, China; IN2P3-CNRS, France; CC-IN2P3, France; SRNSF, Georgia; DFG, Germany; JINR; IBS, Korea; RFBR, Russia; STFC, United Kingdom; and Royal Society, \mbox{United Kingdom}.}
% \acknowledgments{In this section you can acknowledge any support given which is not covered by the author contribution or funding sections. This may include administrative and technical support, or donations in kind (e.g., materials used for experiments).}

\conflictsofinterest{The author declares no conflict of interest.}
% \conflictsofinterest{Declare conflicts of interest or state ``The authors declare no conflict of interest.'' Authors must identify and declare any personal circumstances or interest that may be perceived as inappropriately influencing the representation or interpretation of reported research results. Any role of the funders in the design of the study; in the collection, analyses or interpretation of data; in the writing of the manuscript, or in the decision to publish the results must be declared in this section. If there is no role, please state ``The funders had no role in the design of the study; in the collection, analyses, or interpretation of data; in the writing of the manuscript, or in the decision to publish the~results''.} 

\printendnotes[custom]
%%%%%%%%%%%%%%%%%%%%%%%%%%%%%%%%%%%%%%%%%%
%\end{paracol}    % needed for compiling MDPI style 
%%%%%%%%%%%%%%%%%%%%%%%%%%%%%%%%%%%%%%%%%%
\reftitle{References}


%\externalbibliography{yes}
%\bibliography{your_external_BibTeX_file}

%=====================================
% References, variant B: internal bibliography
%=====================================

\begin{adjustwidth}{-\extralength}{0cm}
%\centering %% If there is a figure in wide page, please release command \centering

\end{adjustwidth}

% If authors have biography, please use the format below
%\section*{Short Biography of Authors}
%\bio
%{\raisebox{-0.35cm}{\includegraphics[width=3.5cm,height=5.3cm,clip,keepaspectratio]{Definitions/author1.pdf}}}
%{\textbf{Firstname Lastname} Biography of first author}
%
%\bio
%{\raisebox{-0.35cm}{\includegraphics[width=3.5cm,height=5.3cm,clip,keepaspectratio]{Definitions/author2.jpg}}}
%{\textbf{Firstname Lastname} Biography of second author}

% The following MDPI journals use author-date citation: Admsci,  Arts, Econometrics, Economies, Genealogy, Humanities, IJFS, Jintelligence, JRFM, Languages, Laws, Literature, Religions, Risks, Social Sciences. For those journals, please follow the formatting guidelines on http://www.mdpi.com/authors/references
% To cite two works by the same author: \citeauthor{ref-journal-1a} (\citeyear{ref-journal-1a}, \citeyear{ref-journal-1b}). This produces: Whittaker (1967, 1975)
% To cite two works by the same author with specific pages: \citeauthor{ref-journal-3a} (\citeyear{ref-journal-3a}, p. 328; \citeyear{ref-journal-3b}, p.475). This produces: Wong (1999, p. 328; 2000, p. 475)

%%%%%%%%%%%%%%%%%%%%%%%%%%%%%%%%%%%%%%%%%%
%% for journal Sci
%\reviewreports{\\
%Reviewer 1 comments and authors’ response\\
%Reviewer 2 comments and authors’ response\\
%Reviewer 3 comments and authors’ response
%}
%%%%%%%%%%%%%%%%%%%%%%%%%%%%%%%%%%%%%%%%%%
\end{document}